\documentclass[preprint]{aastex}
\usepackage{chrisfigcaption}
\usepackage{chrisshortcuts}
\usepackage{graphicx}

\slugcomment{Published in ApJS, 150, 239 [2004]. 
This version incorporates corrections due to accepted Erratum ApJS, 153 [2004]}

\shorttitle{AST/RO: Galactic Center I.}
\shortauthors{Martin et al.}

\begin{document}

\title{The AST/RO Survey of the Galactic Center Region. I. The Inner 3
Degrees}

\author{Christopher L. Martin$^1$, Wilfred M. Walsh$^1$, Kecheng Xiao$^1$,
  Adair P. Lane$^1$,\\
  Christopher K. Walker$^2$, and Antony A. Stark$^1$}
\affil{$^1$ Harvard-Smithsonian Center for Astrophysics, 60 Garden St.,
  MS-12, Cambridge, MA 02138 \\
$^2$ Steward Observatory, University of Arizona, 933 N. Cherry Ave., Tucson, AZ 85721}
\email{cmartin, wwalsh, kxiao, adair, aas@cfa.harvard.edu, cwalker@as.arizona.edu}

\begin{abstract}
  We present fully-sampled maps of 461\ghz{} \co{4}, 807\ghz{}
  \co{7}, and 492\ghz{} \cit{1} emission from the inner 3 degrees of 
  the Galactic Center region taken with the Antarctic Submillimeter
  Telescope and Remote Observatory (AST/RO) in 2001--2002.  The data
  cover $-1\pdeg 3 < \gl < 2\deg$, $-0\pdeg 3 < \gb < 0\pdeg 2$ with
  $0.5\arcmin$ spacing, resulting in spectra in 3 transitions at 
  over 24,000 positions on the sky.  The \co{4} emission is found to
  be essentially coextensive with lower-$J$ transitions of CO. The 
  \co{7} emission is spatially confined to a far smaller region than
  the lower-$J$ CO lines.  The \cit{1} emission has a spatial extent
  similar to the low-$J$ CO emission, but is more diffuse. Bright 
  \co{7} emission is detected in the well-known Galactic Center 
  clouds \sgr{A} and \sgr{B}.  We also detect \co{4} and \co{7}
  absorption from spiral arms in the galactic disk at 
  velocities near 0~\kms{} along the line of sight to the
  Galactic Center. 
  Analyzing our \co{7} and \co{4} data in conjunction with
  $J = 1 \rightarrow 0$ $^{12}{\mathrm{CO}}$ and $^{13}{\mathrm{CO}}$
  data previously observed with the Bell Laboratories 7-m antenna,
  we apply a Large Velocity Gradient (LVG)
  model to estimate the kinetic temperature and density of molecular
  gas in the inner $200\pc$ of the Galactic Center region.
  We show maps of the derived
  distribution of gas density and kinetic temperature as a
  function of position and velocity for the entire region.
  Kinetic temperature was found to decrease from
  relatively high values ($>$70 K) at cloud edges to low values ($<$50 K)
  in the interiors.
  Typical gas pressures in the Galactic Center gas are
  $\hden \cdot \tkin \sim 10^{5.2} \, \mathrm{K \, cm^{-3}}$.
  We present an \glb{} map of molecular hydrogen column density derived
  from our LVG results. 
\end{abstract}

\keywords{Galaxy:center --- Galaxy:kinematics and dynamics --- ISM:atoms --- 
ISM:general --- ISM:molecules \\[0.5cm]
}

\section{Introduction}
\label{sec:intro}

Much has been learned about dense gas in the Galactic
Center region through radio spectroscopy.
Early observations of $F(2 \rightarrow 2)$ OH absorption
\citep{1964:Robin.Gardn.vanDa.Bolto,1964:Golds.Gunde.Penzi.Lille} 
suggested the existence of copious molecular material
within 500\pc{} of the Galactic Center. 
This was confirmed by detection of extensive 
$J = 1 \rightarrow 0$ $^{12}{\mathrm{CO}}$ emission
\citep{1977:Bania,1978:Liszt.Burto}.
Subsequent CO surveys 
\citep{
1987:Bitra,
1988:Stark.Bally.Knapp.Wilso,
1998:Oka.Haseg.Sato.Tsubo,
1997:Bitra.Alvar.Bronf.May} 
have measured this emission with
improving coverage and resolution---these surveys 
show a complex distribution of emission, 
which is chaotic, asymmetric, and non-planar; there are 
hundreds of clouds, shells, arcs, rings, and filaments.
On scales of 100\pc{} to 4\kpc{}, however, the gas
is loosely organized around closed orbits in the
rotating potential of the underlying
stellar bar \citep{1991:Binne.Gerha.Stark.Bally}.
Some CO-emitting gas is bound into clouds and cloud complexes,
and some is sheared by tidal forces into a molecular inter-cloud medium
of a kind not seen elsewhere in the Galaxy \citep{1989:Stark.Bally.Wilso.Pound}.
This diffuse inter-cloud medium appears in absorption
in $F( 2 \rightarrow 2)$ OH \citep{1970:McGee,1970:Robin.McGee},
in ($1_{10} \rightarrow 1_{11}$) H$_2$CO 
% \citep{1972:Scovi.Solom.Thadd}, 
\nocite{1972:Scovi.Solom.Thadd} (Scoville, Solomon, \& Thaddeus 1972),
and in $J = 0 \rightarrow 1 $ HCO$^{+}$ and HCN 
% \citep{1981:Linke.Stark.Frerk}.
\nocite{1981:Linke.Stark.Frerk} (Linke, Stark \& Frerking 1981).
In contrast, the clouds and cloud complexes are dense, as they must be 
to survive in the galactic tide, 
and they appear in spectral lines 
which are tracers of high density 
($\hden{} > 10^4 \, \mathrm{cm}^{-3}$), such 
as NH$_3$ (1, 1) 
% \citep{1981:Guest.Walms.Pauls} 
\nocite{1981:Guest.Walms.Pauls} (G\"{u}sten, Walmsley, \& Pauls 1981)
and CS $J = 2 \rightarrow 1$ \citep{1988:Bally.Stark.Wilso.Henke}.
The large cloud complexes, \sgr{A}, \sgr{B}, and \sgr{C},
are the among the largest molecular cloud complexes in the Galaxy
($M \gsim 10^{6.5} \, \mathrm{M_{\odot}}$).
Such massive clouds must be sinking toward the center
of the galactic gravitational well as a result of dynamical
friction and hydrodynamic effects \citep{1991:Stark.Bally.Gerha.Binne}. 
The deposition of these massive lumps of gas upon
the center could fuel a starburst or an eruption of
the central black hole \citep{1987:Genze.Towne}.

As prelude to further study of the
Galactic Center molecular gas, we would like to determine its
physical state---its temperature and density. 
This involves 
understanding radiative transfer in CO, the primary
tracer of molecular gas.  Also useful is an understanding
of the atomic carbon lines, \cibracket, since those lines trace the more
diffuse molecular regions, where CO is destroyed by UV radiation
but H$_2$ is still present.

The $J = 1 \rightarrow 0$ $^{12}{\mathrm{CO}}$ line is
often optically thick.
Its optical depth can be estimated 
by studying its isotopomers, 
$^{13}{\mathrm{CO}}$ 
and ${\mathrm{C^{18}O}}$.
In the Galactic Center region,
$^{13}{\mathrm{CO}}$ 
is $\sim 24 \times$ less  
abundant than $^{12}{\mathrm{CO}}$
\citep{1980:Penzi,1992:Wilso.Matte}, and
${\mathrm{C^{18}O}}$ is $\sim 250 \times$ less
abundant than
$^{12}{\mathrm{CO}}$
\citep{1981:Penzi}.  Since the radiative
and collisional constants of all the isotopomers are similar,
the ratio of optical depths in their various
spectral lines should simply reflect their
relative abundances. Where the lines are optically
thin, the line brightnesses should be in the same
ratio as the isotopic abundances; where the lines are thick,
deviations from the abundance
ratios are a measure of optical depth.
\citet{
1987:Bally.Stark.Wilso.Henke,
1988:Bally.Stark.Wilso.Henke}
and \citet{1988:Stark.Bally.Knapp.Wilso} 
produced fully-sampled surveys of 
$^{12}{\mathrm{CO}}$
and 
$^{13}{\mathrm{CO}}$ 
in the Galactic Center region.  
They find that the ratio of the
$^{12}{\mathrm{CO}}$ $J = 1 \rightarrow 0$
to
$^{13}{\mathrm{CO}}$ $J = 1 \rightarrow 0$
line brightness temperatures 
($T_{1\rightarrow0}^{12} / T_{1\rightarrow0}^{13}$) 
is typically $10 \pm 2$ in Galactic Center gas that is far
from dense cloud cores.
This indicates much of the Galactic Center
$^{12}{\mathrm{CO}}$ emission is only moderately thick 
($\tau_{1\rightarrow0}^{12} \sim 2$),
especially in comparison to the galactic disk outside 3\kpc{}
radius, where 
$T_{1\rightarrow0}^{12} / T_{1\rightarrow0}^{13} \sim 6$
\citep{1988:Polk.Knapp.Stark.Wilso} is smaller,
even though the isotope ratio 
$^{12}\mathrm{C}/{^{13}\mathrm{C}} \sim 40$ \citep{1980:Penzi}
is larger.
\citet{1982:Heili} and \citet{1998:Dahme.Huett.Wilso.Mauer} 
made surveys in 
${\mathrm{C^{18}O}}$ $J = 1 \rightarrow 0$.  
These show
$^{12}{\mathrm{CO}}$ $J = 1 \rightarrow 0$
to
${\mathrm{C^{18}O}}$ $J = 1 \rightarrow 0$
line brightness temperature ratios 
($T_{1\rightarrow0}^{12} / T_{1\rightarrow0}^{18}$) 
which vary from 40 to over 200, with typical values
near 70, indicating values of $\tau_{1\rightarrow0}^{12}$
which vary from 3 to less than 1,
while the core region of \sgr{B2} shows
$\tau_{1\rightarrow0}^{12} \sim 10$.

Determining the excitation temperature of CO works best
if emission lines from several $J$ levels have been 
measured.  
Lacking such observations, what is often done is 
to use the brightness temperature of the
$^{12}{\mathrm{CO}}$ $J = 1 \rightarrow 0$
line as a lower limit to the excitation temperature
of the $J=1$ state, $T_{\mathrm{ex},J=1}$. 
This estimate can be misleading, because the emission may not
fill the telescope beam, diluting the brightness temperature
and causing it to be many times smaller than 
$T_{\mathrm{ex},J=1} $;
as will be apparent from the data to be presented here, 
this is the usual case for gas in the
Galactic Center region.

Moving up the energy ladder,
\citet{2001:Sawad.Haseg.Handa.Morin} surveyed
the Galactic Center region in
$^{12}{\mathrm{CO}}$ $J = 2 \rightarrow 1$.  They compare their
data to the $J = 1 \rightarrow 0$ data
of \citet{1997:Bitra.Alvar.Bronf.May} and find 
$T_{2\rightarrow1}^{12} / T_{1\rightarrow0}^{12} = 0.96 \pm 0.01$,
with little spatial variation.
%They also observed $^{13}{\mathrm{CO}}$ $J = 2 \rightarrow 1$
%in a strip at $b = 0\deg$, and find 
%$T_{2\rightarrow1}^{12} / T_{2\rightarrow1}^{13} = 10 \pm 1$,
%identical to the value of
%$T_{1\rightarrow0}^{12} / T_{1\rightarrow0}^{13}$ found by
%\citet{1987:Bally.Stark.Wilso.Henke,1988:Bally.Stark.Wilso.Henke}.
%
%Note---there is a real problem with this.  Suppose the 
%ratio in 
%$T_{1\rightarrow0}^{12} / T_{1\rightarrow0}^{13}$ really is
%10, as in the previous paragraph, and this is really a 
%manifestation of moderate optical depth, as we assert.
%Then the 
%T_{2\rightarrow1}^{13} line should have 4 x the optical
%depth of the 
%T_{1\rightarrow0}^{13} line, and the  
%$T_{2\rightarrow1}^{12} / T_{2\rightarrow1}^{13}$
%would then have to be 4x less.  What gives?
What this means is that 
almost all the CO in the Galactic Center region
has low-$J$ states which are
close to local thermodynamic equilibrium (LTE), so that the
excitation temperatures $T_{\mathrm{ex},J}$ of those states are all close to
the kinetic temperature, $T_{\mathrm{kin}}$, and the  
ratio of line brightnesses for transitions between those states
are near unity and therefore independent of $\, T_{\mathrm{kin}}$
\citep[cf.][]{1974:Goldr.Kwan}.
LTE in the low-$J$ states of CO does not occur under all
circumstances in the interstellar medium, but it 
is very common and appears
to be the rule for Galactic Center gas.
For each value of $T_{\mathrm{kin}}$ and $n(\mathrm{H}_2)$, 
there will, however, be some 
value of $J$ above which all higher-$J$ states
fail to be populated, because their Einstein $A$ coefficients
(which increase as $J^3$) are so large that 
the collision rate 
at that value of $T_{\mathrm{kin}}$ and $n(\mathrm{H}_2)$
cannot maintain those states in LTE,
and they must therefore be subthermally excited, i.e.,
$T_{\mathrm{ex},J} <<  T_\mathrm{kin}$.
The brightness temperature of the $J \rightarrow J-1$ line from
those states will be significantly less than that of the 
lower-$J$ states, and the line ratios
$T_{J\rightarrow {J-1}}^{12} / T_{1\rightarrow0}^{12}$ will be
much smaller than unity.
Unlike the low-$J$ states,
the value of the line ratios from those higher-$J$ states
will vary from place to place, depending on 
$T_{\mathrm{kin}}$, $n(\mathrm{H}_2)$, and radiative transfer effects. 

Higher still on the energy ladder,
\citet{2002:Kim.Marti.Stark.Lane}
used the AST/RO telescope to survey a strip at $b = 0\deg$ in 
$^{12}{\mathrm{CO}}$ $J = 4 \rightarrow 3$  and
$^{12}{\mathrm{CO}}$ $J = 7 \rightarrow 6$.  They found
that even the
$T_{4\rightarrow3}^{12} / T_{1\rightarrow0}^{12}$ ratio is
not far from unity and shows
little spatial variation.  
In contrast, the distribution of $J = 7 \rightarrow 6$  
line emission
was found to be markedly different from the lower-$J$ transitions.
Temperatures and densities could therefore be calculated as a 
function of position and velocity using the varying value of
$T_{7\rightarrow6}^{12} / T_{4\rightarrow3}^{12}$
and an estimate 
of $\tau_{1\rightarrow0}^{12}$ from the 
\citet{ 
1987:Bally.Stark.Wilso.Henke,
1988:Bally.Stark.Wilso.Henke} and
\citet{1988:Stark.Bally.Knapp.Wilso}
data.
In the current paper, we extend this work to a fully-sampled
$(\ell , b)$ map of the Galactic Center region, and estimate
kinetic temperature, $T_{\mathrm{kin}}$, and
density, $n(\mathrm{H_2})$, throughout our mapped area. 

\citet{2001:Ojha.Stark.Hsieh.Lane} used AST/RO to make a coarse 
survey of the Galactic Center region in the 
($^3\mathrm{P}_1{\rightarrow} {^3{\mathrm{P}}_0}$) line of 
\cibracket{} at 492\ghz{}.  They find that 
\cibracket{} is distributed approximately like CO, but that the ratio
$T_{\ci} / T_{1\rightarrow0}^{12}$ is smaller in the Galactic
Center region than in the disk outside 3\kpc{}.  This is 
attributable to the smaller average optical depth of the
$^{12}{\mathrm{CO}}$ $J = 1 \rightarrow 0$  in this region.
The absence of strong spatial variations of 
$T_{\mathrm{C\,I}} / T_{1\rightarrow0}^{12}$
across their map indicates that the amounts of CO-emitting
and \cibracket{}-emitting gas are approximately proportionate
across the map.  In the current
paper we extend this work to a more detailed, 
fully-sampled \glb{} map in \cibracket .

In \S\,\ref{sec:obs}, we describe the observatory and the
observations.  
In \S\,\ref{sec:maps}, the full data set in each of the three
observed transitions is presented in the form of spatial--spatial
maps integrated over the full velocity range, velocity-channel 
maps, and spatial--velocity maps at several galactic latitudes.
In \S\,\ref{sec:lvg}, we present a Large Velocity Gradient (LVG)
model of the radiative transfer in CO, which allows a
determination of kinetic temperature, \tkin{}, and 
density, \hden{}.
In \S\,\ref{sec:conc}, we present our conclusions.

\section{Observations}
\label{sec:obs}

%\subsection{Observatory}

The observations were performed during the austral winter seasons of 2001 and 2002
at the Antarctic Submillimeter Telescope and Remote Observatory (AST/RO)
% \citep[AST/RO;][]{2001:Stark.Bally.Balm.Bania}, 
located at 2847\units{m} altitude at the Amundsen-Scott South Pole Station.  
This
site has very low water vapor, high atmospheric stability and a thin
troposphere making it exceptionally good for submillimeter observations
%\citep{1997:Chamb.Lane.Stark, 1998:Lane}. 
\nocite{1997:Chamb.Lane.Stark} \nocite{1998:Lane} (Chamberlin, Lane, \& Stark 1997; Lane 1998).
AST/RO is a 1.7\units{m}
diameter, offset Gregorian telescope capable of observing at wavelengths
between 200\microns{} and 1.3\units{mm}
% \citep{1997:Stark.Chamb.Cheng.Ingal, 2001:Stark.Bally.Balm.Bania}.
\nocite{1997:Stark.Chamb.Cheng.Ingal} \nocite{2001:Stark.Bally.Balm.Bania} (Stark et al. 1997, 2001).
A dual-channel SIS waveguide receiver
\citep{1992:Walke.Kooi.Chan.Leduc,1997:Honin.Haas.Hottg.Jacob}
was used for simultaneous 461--492\ghz{} and 807\ghz{} observations, with
double-sideband noise temperatures of 320--390\units{K} and
1050--1190\units{K}, respectively. Telescope efficiency, $\eta_{\ell}$, estimated
using moon scans, skydips, and measurements of the beam edge taper, was
81\% at 461--492\ghz{} and 71\% at 807\ghz{}. Atmosphere-corrected system
temperatures ranged from 700 to 4000\units{K} at 461--492\ghz{} and 9000 to
75,000\units{K} at 807\ghz{}.

%\subsection{Data Collection}

A multiple position-switching mode was used, with emission-free
reference positions chosen at least $60\arcmin$ from regions of
interest. These reference positions were then shared by a strip of
points at constant galactic latitude which were observed five at a time.
This mapping mode caused each point in the map to be observed for
60\units{s} per pass through the map. In an attempt to
obtain uniform noise over the entire region of interest, subregions were
reimaged as often as required.  

Emission from the \co{4} and
\co{7} lines at 461.041\ghz{} and 806.652\ghz{}, together with the
\cit{1} and \cit{2} lines at 492.262\ghz{} and 809.342\ghz{}, was
imaged over the Galactic Center region $-1\pdeg 3 < \gl < 2\deg$, $-0\pdeg
3 < \gb < 0\pdeg 2$ with $0.5\arcmin$ spacing in \gl{} and \gb{}; i.e.,
a spacing of a half-beamwidth or less. Smaller selected
areas were also observed with longer integration times in the \cit{1}
line.  Maximum pointing errors were no larger than $1\arcmin$, and the 
beam sizes (FWHM) were $103$--$109\arcsec$ at
461--492\ghz{} and $58\arcsec$ at 807\ghz{}
\citep{2001:Stark.Bally.Balm.Bania}.
To facilitate comparison of the various transitions, the data were
regridded onto a $0.25\arcmin$ grid and smoothed 
to a FWHM spatial resolution of $2\arcmin$
with a Gaussian filter function.

Two acousto-optical spectrometers (AOSs; Schieder, Tolls, \& Winnewisser 1989)
\nocite{1989:Schie.Tolls.Winne}
% \citep[AOSs;][]{1989:Schie.Tolls.Winne}
were used as backends. The AOSs had 1.07~\mhz{} resolution and
0.75\ghz{} effective bandwidth, resulting in velocity resolution of 
0.65\kms{} at
461\ghz{} and 0.37\kms{} at 807\ghz{}. 
To facilitate comparison, the data were then smoothed to a 
uniform velocity resolution of 1\kms{}.
The high frequency observations were
made with the CO $J = 7 \rightarrow 6$ line in the 
lower sideband (LSB). Since the
intermediate frequency of the AST/RO system is 1.5\ghz{}, the \cit{2}
line appears in the upper sideband (USB) and is superposed on the observed
LSB spectrum. The local oscillator frequency was chosen so that the nominal
line centers appear separated by 100\kms{} in the double-sideband spectra.
%A third AOS, used for only a few spectra,
%had 0.031MHz resolution and 0.25GHz bandwidth. 
The standard chopper wheel calibration technique was employed, implemented
at AST/RO by way of regular (every few minutes) observations of the sky and
two blackbody loads of known temperature
\citep{2001:Stark.Bally.Balm.Bania}. Atmospheric transmission was monitored
by regular skydips, and known, bright sources were observed every
few hours to further check calibration and pointing. At periodic intervals
and after tuning, the receivers were manually calibrated against a
liquid-nitrogen-temperature load and the two blackbody loads at ambient
temperature and about 100\units{K}. The latter process also corrects for
the dark current of the AOS optical CCDs. The intensity calibration
errors became as large as $\pm15$\% during poor weather periods.

%\subsection{Data Reduction}
%\label{subsec:reduction}

Once taken, the data in this survey were reduced using the COMB data
reduction package.  After elimination of scans deemed faulty for
various instrumental or weather-related reasons ($\lsim 2\%$ of the
total dataset), linear baselines were removed from the spectra in all
species by excluding regions where the \co{1} spectra showed
emission greater than $T_{\mathrm{A}}^\star \geq 1\units{K}$.  This
allowed known emission in the Galactic Center region to be readily
excluded from the baseline fitting procedure and was generally
sufficient.  In a few cases, usually due to higher than average
\trms{} for a given reduced spectrum, this method fails and
artifacts (e.g., vertical lines in the longitude--velocity maps) appear.  

While the original intent was to make \trms{} as uniform as
possible across the entire map, this was not always possible.
For the \co{4} transition, \trms{} in 1\kms{} wide channels 
with $2\arcmin$ spatial smoothing is on average $\lsim
0.3 \units{K}$ except in the region $1\pdeg 8 > \gl > 1\pdeg 5$ where
$\trms \lsim 0.8\units{K}$.  The \cit{1} transition has
$\trms \lsim 0.5\units{K}$ in 1\kms{} channels 
for the central region of $1\pdeg 0 >
\gl > -0\pdeg 5$, $\trms \lsim 1.0\units{K}$ for $\gl > 1\pdeg 0$,
and $\trms \lsim 2\units{K}$ for $\gl < -0\pdeg 5$.  Finally, for
\co{7} (ignoring the occasional baseline feature), 
$\trms \lsim 0.8\units{K}$ in 1\kms{} channels 
for $\gl > 0.7$ and $-0.5 > \gl > -0.8$,
and $\trms \lsim 2\units{K}$ elsewhere.

\section{Data Presentation}
\label{sec:maps}

Sample spectra at the respective positions of peak \co{7} emission
toward the \sgr{A} (\gl = $0\pdeg 00$, \gb = $-0\pdeg 07$), \sgr{B}
(\gl = $0\pdeg 66$, \gb = $-0\pdeg 05$), \sgr{C} (\gl = $-0\pdeg 45$,
\gb = $-0\pdeg 20$) (\co{4} emission peak), and $\gl \simeq 1\pdeg3$ (\gl = $1\pdeg 25$,
\gb = $-0\pdeg 05$) molecular complexes are shown in
Fig.~\ref{spectra}.  The \co{4}, \co{7}, and \cit{1} data are from the
AST/RO survey, while the \co{1}, $^{13}$\co{1}, and \cs{2} data are
from the Bell Laboratories (BL) 7-m telescope
\citep{1988:Stark.Bally.Knapp.Wilso, 1987:Bally.Stark.Wilso.Henke,
1988:Bally.Stark.Wilso.Henke}.  All of the spectra are from datacubes
smoothed to $2\arcmin$ resolution.

The \co{4} profile toward Sgr~B resembles the \co{1} profile, with similar
linewidth and a prominent self-absorption feature at $\vlsr = 60~\kms$.
The total velocity extent is somewhat smaller in \co{4} on the negative
velocity side.  In contrast, the \cit{1} and \co{7} lines, as well
as the $^{13}$CO, show peak emission at the self-absorption velocity,
suggesting these lines are less optically thick.  It is important to note
that the strong feature at negative velocities in the \co{7} spectrum toward Sgr~B
is due to superposed emission in the 809\ghz{} ($^3$P$_2 \rightarrow ^3$P$_1$) line of 
\cibracket{} in the image sideband.

Fig.~\ref{lbmap_integrated} presents spatial--spatial \glb{}
maps integrated over velocity for the three transitions observed with
AST/RO\footnote{Complete FITS datacubes of the three AST/RO datasets are archived with the electronic edition of this paper.  As new data becomes available updated versions will be posted at \anchor{http://cfa-www.harvard.edu/~adair/AST_RO/abc.html}{http://cfa-www.harvard.edu/$\sim$adair/AST\_{}RO/abc.html}.} and, for comparison, the three transitions observed at the BL
7-m. 
% \cit{1}   v_min=-90  v_max=150
% \co{7}    v_min=-30  v_max=120
% \co{4}    v_min=-150 v_max=150
% \co{1}    v_min=-150 v_max=150
% 13 \co{1} v_min=-150 v_max=150
% \cs{2}    v_min=-150 v_max=150
All six maps have been smoothed to the same $2\arcmin$ spatial resolution. 
The regions with  $\gl > 0\pdeg 9$ and 
$\gl < -0\pdeg 5$ in the $^{13}$\co{1} map were observed with sparser
sampling than the rest of the map \citep{1987:Bally.Stark.Wilso.Henke}.
The most striking result is that \co{4} emission in the Galactic
Center region is essentially coextensive with the emission from
the lower $J$ transitions of CO. This contrasts sharply with the outer
Galaxy where \co{4} emission is rather less extensive than \co{1}.
In all six maps, the brightest emission occurs primarily at negative
latitudes. Four major cloud complexes are seen in the \co{1},
\co{4}, and $^{13}$\co{1} maps,
from left to right: the complex at $\gl \simeq 1\pdeg3$, the \sgrb{} complex
near $\gl \simeq 0\pdeg 7$, the \sgr{A} cloud near $\gl \simeq 0\pdeg 0$,
and the \sgr{C} cloud near ($\gl \simeq -0\pdeg 45$,
$\gb \simeq -0\pdeg 2$). As
noted by \citet{2002:Kim.Marti.Stark.Lane}, the \co{7} emission is
much more spatially confined than the lower-J CO transitions. In
contrast, the \cibracket{} emission is comparable in spatial extent to
the low-J CO emission, but its distribution appears somewhat more
diffuse (less peaked). The \sgr{C} cloud is much less prominent in
the \cibracket{} map than in the other five transitions. The noise in
the \cibracket{} map at $\gl < -0\pdeg 7$ is greater than that in the
rest of the map due to shorter integration times.

Figs.~\ref{lbmap_part_461_01}--\ref{lbmap_part_807_01} present
spatial--spatial maps in successive 10\kms{} wide velocity ranges 
in the \co{4}, \cit{1}, and \co{7} lines, respectively. 
For each spectral line, 
velocity channels centered at velocities from $-165 \kms$ to $+185 \kms$ 
(see upper left corner of each map) are displayed. In each map, the
integrated intensity has been divided by 10\kms{} so that the intensity
color scale is a reasonably good indicator of corrected average antenna 
temperature, $\langle T_\mathrm{A}^\star \rangle$.

Spatial--velocity maps at 12 different values of galactic latitude
are shown in Figs.~\ref{lvmap_461_01}--\ref{lvmap_807_01} for 
\co{4}, \cit{1}, and \co{7}, respectively. All major structures in
the galactic core 
can be identified in these maps, including \sgr{A}, \sgr{B}, 
\sgr{C}, and the $\gl \simeq 1\pdeg 3$ complex. The 300\pc{} molecular
ring shows up prominently in both the \cibracket~ and \co{4} (\gl -- $v$)
maps, running in a straight, continuous line from 
($\gl \sim 2\deg, \vlsr \sim 220 \kms$) to 
($\gl \sim -0\pdeg 9, \vlsr \sim 120 \kms$), and from 
($\gl \sim 1\pdeg 5, \vlsr \sim -20 \kms$) to 
($\gl \sim -0\pdeg 9, \vlsr \sim -180 \kms$). There is little or no 
evidence that the \co{7} line is excited into emission in the
300\pc{} ring. Also absent in the \co{7} line is any trace 
of the 3\kpc{} arm, which
is prominent in \co{1} emission. The 3\kpc{} arm must contain little
dense, warm gas. Foreground absorption by spiral arms in the galactic 
disk is seen at velocities near 0\kms{} in \co{4} and \cit{1}.
A hint of this absorption feature is seen in \co{7}.

\section{LVG Model}
\label{sec:lvg}

The large velocity gradient (LVG) approximation 
\citep{1974:Goldr.Kwan}
simplifies radiative transfer analysis of molecular lines.  
Imagine a small volume within a molecular cloud, in which a
molecule emits a photon.  The volume is sufficiently
small that all velocities within it are thermal, and 
temperature and density are constant.  It
is surrounded by other small volumes which have similar 
temperature and density, and whose internal velocities are
also thermal, but the velocities of these other volumes are
different from that of the volume which emitted the photon because
of velocity gradients within the molecular cloud.
In LVG, it is assumed that if the photon is going to be absorbed,
rather than escaping from the cloud, it can only be absorbed
in nearby volumes, those volumes whose velocities are close
to the velocity of the emitting volume.  In this
case `close' means close enough that the thermal linewidths overlap. 
As the emitted photon travels away from its point of origin, it
passes through nearby volumes where it has a chance to be
absorbed because the thermal motion of some of the molecules causes
them to have the same velocity as the emitting molecule.
At some distance from the point of emission, this will no
longer be true because of velocity gradients, and in LVG
it is assumed that the photon then escapes from the cloud.
The velocity gradient is `large' in the sense that the region
of possible absorption is small compared to the cloud as a 
whole, so it can be assumed that in effect the photon was absorbed at
its point of origin.  This achieves a great simplification
in making the radiative transfer problem entirely `local',
while allowing for the possibility of absorption.
The probability that a photon will escape from the cloud
and become observable
then depends only on the local physical and 
chemical properties of the emitting gas and on the value of
the velocity gradient.  The emitting cloud can then be modeled
using only a few parameters, yet the radiative
transfer is realistic enough to compare to observations.  The
parameters of the model can then be adjusted to fit the data, and an
estimate of those parameters is thereby achieved.
\citet{1997:Ossen} has shown that such estimates are robust,
in the sense that the parameter values derived are often approximately
correct, even in circumstances where the assumptions of
the LVG approximation are violated.

We will use this LVG methodology to estimate the kinetic 
temperature, \tkin, and the number density of molecular 
hydrogen, \hden{}, throughout the Galactic Center region. 
Due to the high velocity dispersions characteristic of the Galactic
Center, the LVG approximation is most likely valid over much of the
mapped region.  
The LVG approximation does not apply 
to some of our data: the foreground absorption by spiral
arms is clearly not local to the emitting gas in the 
\sgr{A} cloud.  It should be kept in mind that the LVG 
approximation is an {\em Ansatz} which allows us to estimate the
physical properties of the Galactic Center gas in what otherwise 
would be an untenably complex modeling problem.

Our cloud model was developed by M. Yan and S. Kim. 
It has plane-parallel cloud geometry. 
It uses CO collisional rates determined by \citet{1995:Turne}
and uses newly-derived values 
for the H$_2$ ortho-to-para ratio ($\approx$ 2)
\citep{2000:Rodri.Marti.Vince} and for 
the collisional quenching rate of CO by H$_2$ impact 
% \citep{2002:Balak.Yan.Dalga}.
\nocite{2002:Balak.Yan.Dalga} (Balakrishnan, Yan, \& Dalgarno 2002).
The model has two input parameters: 
the ratio of $^{12}$CO to $^{13}$CO 
abundance, and the ratio $X(CO) / \nabla V$, 
where $X(CO)$ is the fractional 
CO abundance parameter and $\nabla V$ is the velocity gradient.
As discussed in \S 1, the abundance 
ratio $^{12}$CO/$^{13}$CO is 24 in the Galactic 
Center region 
\citep{1980:Penzi,1990:Lange.Penzi,1992:Wilso.Matte,1993:Lange.Penzi}.  
We will use a value 
$X(CO) / \nabla V= 10^{-4.5} \, {\mathrm{\pc \, km^{-1} \, s}}$, assuming 
that the $^{12}$CO/H$_2$ ratio is 10$^{-4}$ and the velocity gradient 
within the Galactic Center gas is a uniform 3 $\kms\units[-1]{pc}$. 
\citet{1998:Dahme.Huett.Wilso.Mauer} estimated that
$\nabla V$ is 
$3 \, \kms\units[-1]{pc}$ 
to
$6 \, \kms\units[-1]{pc}$, and indeed these are typical slopes of
position-velocity features in Galactic Center maps.  There is,
however, no reason to suppose that a single value of $\nabla V$
applies throughout the Galactic Center region, or even that $\nabla V$
is constant within a single cloud.  This is a weak point in the
analysis, because in the LVG analysis
$ n(\mathrm{H}_2) \propto (\nabla V)^{0.6}$, when all other parameters
are held fixed.

\begin{figure}[H]
\plotone{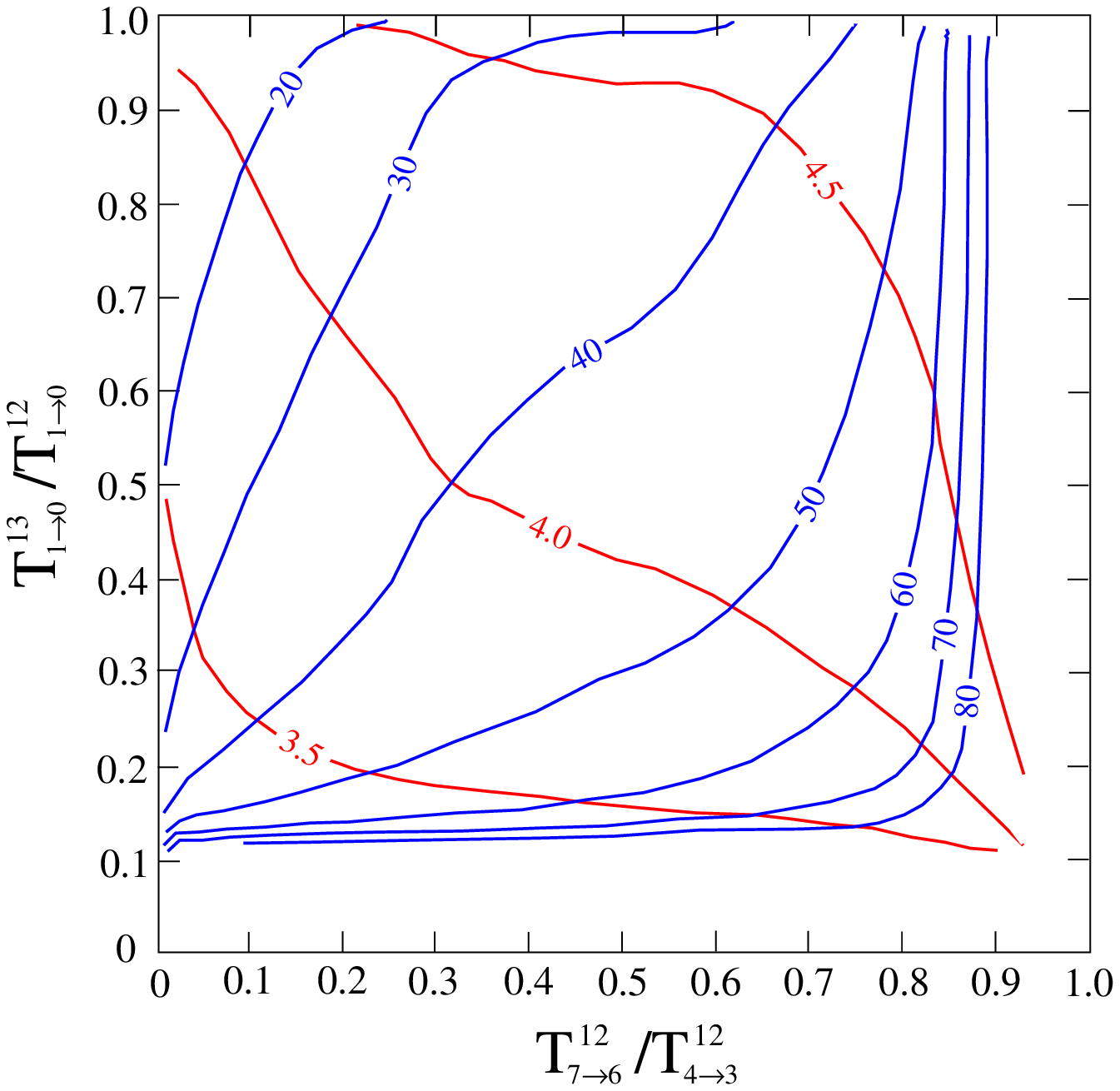}
\caption{Approximate representation of the relation
between the line ratios and \tkin{} (blue curves, units are K) 
and \hden{} (red curves, units are log[$\hden /1.0 \, \mathrm{cm^{-3}}]$) 
generated by our LVG model,
which uses an abundance ratio $^{12}$CO/$^{13}$CO $= 24$ and
$X({\mathrm{CO}}) / \nabla V = 10^{-4.5} \, {\mathrm{\pc \, km^{-1} \, s}}$.
Instabilities in the generation of
these functional relationships have been smoothed by hand.}
\label{lvg}
\end{figure}

For each observed point, we take the brightness temperature ratios 
$T_{7\rightarrow6}^{12} / T_{4\rightarrow3}^{12}$
and
$T_{1\rightarrow0}^{13} / T_{1\rightarrow0}^{12}$,
using the same methodology
as \citet{2002:Kim.Marti.Stark.Lane},
to determine \tkin{} and \hden{}.  
Fig.~\ref{lvg} is a smoothed representation of the relationships 
generated by our LVG model.  Note that our model, using the
$T_{7\rightarrow6}^{12} / T_{4\rightarrow3}^{12}$ ratio,
is particularly sensitive to variations in density and temperature
near 
$\tkin \sim 40 \mathrm{K}$ and
$\hden \sim 10^4 \, \mathrm{cm^{-3}}$, and that its range of
validity extends a factor of $\sim 20$ in density and
$\sim 4$ in temperature around these values.
Much of the Galactic Center gas falls within this
range; the gas which does not could be studied using
various other transitions of the CO isotopomers, or
using other species such as \cibracket \  and CS.

Fig.~\ref{lvmap_tk_02} shows the
output of our LVG analysis for \tkin{} as spatial--velocity maps for 6
different values of galactic latitude, while Fig.~\ref{lvmap_h2den_02}
does the same for \hden{}.  White regions on the maps indicate areas
where either the LVG inversion did not converge or where there is
missing data from the maps of the four transitions used.  
The LVG model strains to fit in regions where
$T_{1\rightarrow0}^{13} / T_{1\rightarrow0}^{12}$ is so small it approaches
the value of the isotope ratio ($1/24$).
Where $T_{1\rightarrow0}^{13} / T_{1\rightarrow0}^{12} < 0.1$, that is,
where the $J = 1 \rightarrow 0$ CO line is less optically thick,
the model assigns high kinetic temperatures
(cf. the bottom of Fig.~\ref{lvg}).
%Moving inward from low density regions, where \hden \, starts low, 
\tkin \, starts high in the lower density regions.
This is physically reasonable since the edges of
these complexes are likely to be of lower density and externally
heated.  
Moving into the interior of the complexes, we find a
surprisingly uniform temperature of $\tkin \sim 50$ K and density
of $\hden \sim 10^{3.5} \, \mathrm{cm^{-3}}$. This indicates a typical pressure 
$\hden \cdot \tkin \sim 10^{5.2} \, \mathrm{K \, cm^{-3}}$. 
Exceptional are the
density peak at \sgr{B} and the foreground material along $\vlsr
\simeq 0\kms$.  The breakdown of LVG assumptions in the foreground
absorbing material renders the \tkin{} and \hden{} results suspect
if not outright invalid near the absorbing material.

Fig.~\ref{lbmap_lvg} shows the estimated column density as determined
by integrating the LVG spatial density and dividing by $\nabla V$.
The region $-60 < \vlsr < 20 \kms$ is excluded in order to avoid
contamination by the foreground material for which the LVG analysis is
invalid; the actual total column density should therefore be somewhat
larger than the
value shown here.  
Also, there are a few places where the LVG model fails to converge even 
though the lines are strong and the density is presumably high---the density
in those spots does not contribute to our estimate of the column
density.
There is an overall factor of order unity uncertainty in the scaling 
of this map due to likely errors in our assumed value of $\nabla V$.
The map clearly shows the
density enhancement toward \sgr{A} and \sgr{B} and the relatively
high density of the material surrounding and connecting them.
In the closed-orbit paradigm of 
\citet{1991:Binne.Gerha.Stark.Bally}, this material is in
$x_2$ orbits.

\section{Conclusions}
\label{sec:conc}

\begin{enumerate}
\item{We have mapped the inner $3\deg$ of the Galaxy in
  461\ghz{} \co{4}, 807\ghz{} \co{7}, and 492\ghz{} \cit{1} emission.
  \sgr{A}, \sgr{B}, \sgr{C}, and the 300\pc{} molecular ring are easily
  identified.
}
\item{The \co{4} emission is found to be essentially coextensive with
  lower-$J$ transitions of CO.  The \co{7} emission is spatially
  confined to a far smaller region than the lower-$J$ CO lines.
  The \cit{1} emission has a spatial extent similar to the low-$J$
  CO emission, but is more diffuse.
}
\item{Consistent with \citet{2002:Kim.Marti.Stark.Lane}, we find the
$T_{4\rightarrow3}^{12} / T_{1\rightarrow0}^{12}$ line ratio is
approximately constant and not far from unity over much of the 
mapped region.  In contrast, the
$T_{7\rightarrow6}^{12} / T_{4\rightarrow3}^{12}$ line ratio is
found to vary significantly.
}
\item{For each observed point, 
$T_{7\rightarrow6}^{12} / T_{4\rightarrow3}^{12}$ line ratios,
together with
$T_{1\rightarrow0}^{13} / T_{1\rightarrow0}^{12}$ line ratios,
were used to estimate kinetic temperatures and molecular hydrogen 
volume densities. Kinetic temperature was found to decrease from
relatively high values ($>$70 K) at cloud edges to low values ($<$50 K)
in the interiors.  Molecular hydrogen densities, \hden{}, ranged up
to the limit of our ability to determine via our LVG analysis,
$\sim 10^{4.5} \, \cm[-3]$. 
}
\item{Typical gas pressures in the Galactic Center gas are
$\hden \cdot \tkin \sim 10^{5.2} \, \mathrm{K \, cm^{-3}}$,
while typical virial pressures are  
$\hden \cdot T_{\mathrm{virial}} \sim 10^{6.8} \, \mathrm{K \, cm^{-3}}$.
These values can be compared to 
the typical gas pressures in molecular clouds near the Sun
$\sim 10^{3.4} \, \mathrm{K \, cm^{-3}}$, 
the typical virial pressure in
molecular clouds near the Sun
$\sim 10^{5} \, \mathrm{K \, cm^{-3}}$, and
the ambient pressure of the interstellar medium 
near the Sun $\sim 10^{4} \, \mathrm{K \, cm^{-3}}$
\citep{1990:Dicke.Lockm}.
}
\end{enumerate}

\acknowledgments
\label{sec:ack}

We thank the receiver group at the U. of Arizona for their assistance; 
R. Schieder, J. Stutzki, and colleagues at U. K\"{o}ln for their AOSs; 
J. Kooi and R. Chamberlin of Caltech, G. Wright of PacketStorm
Communications,
and K. Jacobs of U. K\"{o}ln 
for their work on the instrumentation; 
M. Yan and S. Kim for the use of
their LVG code; and J. Carlstrom for comments on the manuscript.
This research was supported in part by the National Science 
Foundation under a cooperative agreement with the Center for 
Astrophysical Research in Antarctica (CARA), grant number 
NSF OPP 89-20223. CARA is a National Science Foundation 
Science and Technology Center. Support was also provided
by NSF grant number OPP-0126090.

%\appendix

%\section{Appendicial material}

\bibliography{galcen}
\bibliographystyle{apj}

\clearpage

\chrisfigcaption{spectra}{Spectra toward the respective positions of
  peak \co{7} emission in the \sgr{A} (\gl = $0\pdeg 00$, \gb =
  $-0\pdeg 07$), \sgr{B} (\gl = $0\pdeg 66$, \gb = $-0\pdeg 05$),
  \sgr{C} (\gl = $-0\pdeg 45$, \gb = $-0\pdeg 20$) (\co{4} emission peak)`, and $\gl \simeq
  1\pdeg3$ (\gl = $1\pdeg 25$, \gb = $-0\pdeg 05$) clouds (as indicated
  at lower right in each frame) in 6 different transitions, as indicated
  by the color identifications at upper left.  The 461~GHz \co{4},
  807~GHz \co{7}, and 492~GHz \cit{1} data are from the AST/RO survey
  ({\it this paper}), and the 115~GHz \co{1}, 110~GHz $^{13}$\co{1},
  and 98~GHz \cs{2} data are from the Bell Laboratories 7-m telescope
  \citep{1988:Stark.Bally.Knapp.Wilso, 1987:Bally.Stark.Wilso.Henke,
  1988:Bally.Stark.Wilso.Henke}.  }{1}

\chrisfigcaption{lbmap_integrated}{Spatial--spatial \glb{}
  integrated intensity maps for the 3 transitions observed with
  AST/RO ({\it top 3 panels}) and, for comparison, the 3 transitions 
  observed at the BL 7-m \citep{1988:Stark.Bally.Knapp.Wilso,
  1987:Bally.Stark.Wilso.Henke,
  1988:Bally.Stark.Wilso.Henke} ({\it bottom 3 panels}). Transitions
  are identified at left on each panel. 
  The emission is integrated over all velocities
  where data are available.  These values of ($v_{min}$, $v_{max}$)
  are: \cibracket{}, ($-90, 150$); \co[s]{7}, ($-30, 120$); \co[s]{4}, ($-150, 150$);
  \co[s]{1}, ($-150, 150$); $^{13}$\co[s]{1}, ($-150, 150$); \cs[s]{2}, ($-150,
  150$). All 6 maps have been smoothed to the same $2\arcmin$ resolution.  This figure has been corrected as noted in the erratum.}{2}

\chrisfigcaption{lbmap_part_461_01}{False color velocity-channel maps
  for the \co{4} transition toward the Galactic Center.  Each subpanel
  shows a 10\kms{} velocity bin, centered at the velocity shown in
  the upper lefthand corner.  The integrated emission in $\units{K}\,\kms$ has
  been divided by the 10\kms{} width of the bin so that the color scale
  at right indicates $\langle T_{\mathrm{A}}^\star\rangle$.
  This figure has been corrected as noted in the erratum.}{3a}
\chrisfigcaptioncontinued{lbmap_part_461_02}{3b}
\chrisfigcaptioncontinued{lbmap_part_461_03}{3c}

\chrisfigcaption{lbmap_part_492_01}{False color velocity-channel maps
  for the \cit{1} transition toward the Galactic Center.  Each subpanel
  shows a 10\kms{} velocity bin, centered at the velocity shown in
  the upper lefthand corner.  The integrated emission in $\units{K}\,\kms$ has
  been divided by the 10\kms{} width of the bin so that the color scale
  at right indicates $\langle T_{\mathrm{A}}^\star\rangle$.
  This figure has been corrected as noted in the erratum.}{4a}
\chrisfigcaptioncontinued{lbmap_part_492_02}{4b}
\chrisfigcaptioncontinued{lbmap_part_492_03}{4c}

\chrisfigcaption{lbmap_part_807_01}{False color velocity-channel maps
  for the \co{7} transition toward the Galactic Center.  Each subpanel
  shows a 10\kms{} velocity bin, centered at the velocity shown in
  the upper lefthand corner.  The integrated emission in $\units{K}\,\kms$ has
  been divided by the 10\kms{} width of the bin so that the color scale
  at right indicates $\langle T_{\mathrm{A}}^\star\rangle$.
  This figure has been corrected as noted in the erratum.}{5a}
\chrisfigcaptioncontinued{lbmap_part_807_02}{5b}
\chrisfigcaptioncontinued{lbmap_part_807_03}{5c}

\chrisfigcaption{lvmap_461_01}{False color longitude--velocity maps of 
  \co{4} emission toward the Galactic Center.  Each of the 12 panels
  displays emission at a different value of galactic latitude, indicated
  in the lower left corner of each panel.
}{6a}
\chrisfigcaptioncontinued{lvmap_461_02}{6b}
\chrisfigcaptioncontinued{lvmap_461_03}{6c}
\chrisfigcaptioncontinued{lvmap_461_04}{6d}

\chrisfigcaption{lvmap_492_01}{False color longitude--velocity maps of
  \cit{1} emission toward the Galactic Center.  Each of the 12 panels
  displays emission at a different value of galactic latitude, indicated
  in the lower left corner of each panel.
}{7a}
\chrisfigcaptioncontinued{lvmap_492_02}{7b}
\chrisfigcaptioncontinued{lvmap_492_03}{7c}
\chrisfigcaptioncontinued{lvmap_492_04}{7d}

\chrisfigcaption{lvmap_807_01}{False color longitude--velocity maps of
  \co{7} emission toward the Galactic Center.  Each of the 12 panels
  displays emission at a different value of galactic latitude, indicated
  in the lower left corner of each panel.
}{8a}
\chrisfigcaptioncontinued{lvmap_807_02}{8b}
\chrisfigcaptioncontinued{lvmap_807_03}{8c}
\chrisfigcaptioncontinued{lvmap_807_04}{8d}

\chrisfigcaption{lvmap_tk_02}{False color longitude--velocity maps of
  \tkin{} as determined by the LVG model described in the text.  Regions
  in white indicate areas where either spectral line data are not
  available or the LVG model did not converge.  Each of the 6
  panels displays \tkin{} at a different value of galactic latitude,
  indicated in the lower left corner of each panel.}{10a}
\chrisfigcaptioncontinued{lvmap_tk_03}{10b}

\chrisfigcaption{lvmap_h2den_02}{False color longitude--velocity maps of
  $\log\hden$ as determined by the LVG model described in the text.  Regions
  in white indicate areas where either spectral line data are not
  available or the LVG model did not converge.  Each of the 6
  panels displays \hden{} at a different value of galactic latitude,
  indicated in the lower left corner of each panel.}{11a}
\chrisfigcaptioncontinued{lvmap_h2den_03}{11b}

\chrisfigcaption{lbmap_lvg}{False color velocity-channel map of
  $\log[\int \hden dv/\nabla V]$ as determined by the LVG model
  described in the text.  \hden{} is integrated over the ranges $-150
  < \vlsr < -60 \kms$ and $20 < \vlsr < 150 \kms$ in order to avoid
  contamination by the foreground material for which the LVG analysis
  is invalid.  This value is then divided by $\nabla V$ in order to
  make a map comparable to the expected column density in units of
  \cm[-2].}{12}

\end{document}